\documentclass[11pt,a4paper]{article}
\usepackage{amssymb}
\usepackage{times,cite,epsfig}
\newcommand{\be}{\begin{eqnarray}}
\newcommand{\ee}{\end{eqnarray}}
\newcommand{\bez}{\begin{eqnarray*}}
\newcommand{\eez}{\end{eqnarray*}}
\newcommand{\pa}{\partial}
\newcommand{\la}{\lambda}

\newcommand{\A}{\mathcal{A}}
\newcommand{\res}{\mathit{res}}

\newcommand{\tpa}{\tilde{\pa}}
\newcommand{\tE}{\tilde{E}}
\newcommand{\tq}{\tilde{q}}
\newcommand{\tir}{\tilde{r}}

\renewcommand{\d}{\mathrm{d}}

\title{\bf From AKNS to derivative NLS hierarchies \\
   via deformations of associative products}

\author{Aristophanes Dimakis \\
 Department of Financial and Management Engineering, \\
 University of the Aegean, 31 Fostini Str., GR-82100 Chios, Greece \\
 dimakis@aegean.gr
          \and
 Folkert M\"uller-Hoissen \\ Max-Planck-Institute for Dynamics and Self-Organization \\
 Bunsenstrasse 10, D-37073 G\"ottingen, Germany \\
 folkert.mueller-hoissen@ds.mpg.de }

\date{}

\begin{document}

\renewcommand{\theequation} {\arabic{section}.\arabic{equation}}

\newtheorem{theorem}{Theorem}[section]
\newtheorem{lemma}{Lemma}[section]
\newtheorem{proposition}{Proposition}[section]
\newtheorem{definition}{Definition}[section]

\maketitle

\begin{abstract}
Using deformations of associative products, derivative 
nonlinear Schr\"odinger (DNLS) hierarchies are recovered as AKNS-type hierarchies. 
Since the latter can also be formulated as Gelfand-Dickey-type Lax hierarchies, 
a recently developed method to obtain `functional representations' 
can be applied. We actually consider hierarchies with dependent variables 
in any (possibly noncommutative) associative algebra, e.g., an algebra of matrices 
of functions. This also covers the case of hierarchies of coupled 
derivative NLS equations. 
\end{abstract}

\section{Introduction}
\label{section:Intro}
\setcounter{equation}{0} 
An AKNS hierarchy (see \cite{AKNS73,AKNS74,FNR83,Wils81,Drin+Soko84,Dick03,DMH04hier}, 
for example) is given by (a multi-component generalization of)
\be
    \pa_{t_n}(V) = [ (\zeta^n \, V)_{\geq k} , V]   \qquad \quad n=1,2, \ldots
            \label{AKNS-Vhier}
\ee
with independent variables $t_n$, an indeterminate $\zeta$, a formal power 
series $V$ in $\zeta^{-1}$ with coefficients in a Lie algebra $\cal{G}$, 
and a projection $(\, )_{\geq k}$ to terms containing only powers $\geq k$ 
of $\zeta$, where $k=0$ or $k=1$. The Lie bracket used in (\ref{AKNS-Vhier}) 
may depend nontrivially on $\zeta$. Integrable models obtained via such 
deformations have been studied in \cite{Golu+Soko02,Golu+Soko04,Golu+Soko05}
and \cite{Skry02,Skry04a,Skry04b,Skry05} (see also the references in these papers).
We demonstrate that in this way one can also recover hierarchies
associated with derivative nonlinear Schr\"odinger equations (DNLS). 
Instead of working with deformations of a Lie algebra, we consider 
deformations of an associative algebra (which then, via the commutator, 
induce a Lie algebra structure). Although this somewhat restricts the 
possibilities, it allows some technical steps which greatly simplify the 
analysis (see also the remark in section~\ref{section:DNLS&AKNS}). 
The problem of deformations of associative products also arises in 
the context of compatible Poisson structures, which is of relevance 
in the theory of integrable systems too (see \cite{CGM00,Odes+Soko05,Odes+Soko06}, 
in particular). 
\vskip.2cm

In this work we derive moreover \emph{functional representations} of DNLS hierarchies. 
These are generating equations, dependent on auxiliary parameters, 
which determine the hierarchy equations directly
in terms of the relevant dependent variables (see also
\cite{Nijh+Cape90,Bogd+Kono98,Bogd99,BKM03,Bogd+Kono05,Veks98,Veks02,Prit+Veks02,Veks04}).
We make use of a method developed in \cite{DMH06func}, 
which will be recalled in section~\ref{ssection:prel}. 
Here it is of relevance that AKNS hierarchies can be reformulated as 
`Gelfand-Dickey-type' hierarchies (see (\ref{GD-hier})). 
\vskip.2cm

We actually derive functional representations for hierarchies 
with dependent variables in any (possibly noncommutative) associative algebra 
(see also \cite{Olve+Soko98,Olve+Soko98assoc,Kupe00}, for example). 
Specialization to matrix algebras then covers cases of coupled systems of 
equations. These are also a possible source of new integrable equations. 
A further motivation is provided by the operator method 
\cite{March88,Carl+Schi99,Carl+Schi00} which associates with a (scalar) 
nonlinear equation an operator version and then a suitable map from 
solutions of the latter to solutions of the former. For this method it is 
prerequisite, of course, to find an `integrable' generalization of the respective 
equation with dependent variable in a noncommutative associative algebra. 
\vskip.2cm

Section~\ref{section:DNLS&AKNS} shows how derivative NLS hierarchies 
can be expressed as AKNS-type hierarchies by taking the possibility of 
deformations in the sense mentioned above into account.
In subsections we obtain matrix versions of hierarchies associated with
three variants (Chen-Lee-Liu, Gerdjikov-Ivanov, and Kaup-Newell) 
of DNLS equations in this way. 
Section~\ref{section:functional} recalls some results from \cite{DMH06func} 
which are then used to derive functional representations of the DNLS hierarchies.
Section~\ref{section:concl} contains some concluding remarks.
Appendix~A provides elementary material on deformations 
of products, as used in the main part of this work. 
As outlined in appendix~B, in cases where it is possible to `undeform' the product, contact with other formulations of the respective hierarchies can be established.

\section{Derivative NLS hierarchies as AKNS-type hierarchies}
\label{section:DNLS&AKNS}
\setcounter{equation}{0}
Let $(\A,\bullet)$ be an associative algebra (over a field $\mathbb{K}$ of characteristic 
zero, typically $\mathbb{R}$ or $\mathbb{C}$) with unit $J$. 
The product $\bullet$ trivially extends to the algebra $\hat{\A} := \A[\zeta,\zeta^{-1}]]$ 
of polynomials in an indeterminate $\zeta$ and formal power series in $\zeta^{-1}$ 
(with coefficients in $\A$). 
In the following we consider an AKNS-type hierarchy
\be
    \pa_{t_n}(V) = [ (\zeta^n \, V)_{\geq k} , V]_{\bullet}  
       = - [ (\zeta^n \, V)_{<k} , V]_{\bullet} 
    \qquad \quad n=1,2, \ldots \, ,
            \label{AKNS-Vhier2}
\ee
where $[X,Y]_{\bullet} := X \bullet Y - Y \bullet X$. 
Without restriction of generality we may take $V$ of the form
\be
    V = u_0 + u_1 \, \zeta^{-1} + u_2 \, \zeta^{-2} 
           + u_3 \, \zeta^{-3} + \ldots      \label{V-expansion}
\ee
with $u_n \in \A$. Note that (\ref{AKNS-Vhier2}) is invariant under an additive shift 
of $V$ by some constant element in the center of $\hat{\A}$, e.g., any multiple of 
the unit element. As a consequence, there is a certain arbitrariness in the 
choice of $u_0$ in (\ref{V-expansion}). For example, the original AKNS hierarchy (cf.  
\cite{FNR83}, for example) is obtained with an algebra of $2 \times 2$ matrices and 
$u_0 = \mbox{diag}(1,-1)$ (up to multiplication by $-\imath$), 
but we may choose $u_0 = \mbox{diag}(1,0)$ as well. In this work we actually 
concentrate on examples based on algebras of $2 \times 2$ matrices, but with 
entries in an associative and typically noncommutative algebra.\footnote{In the 
case of $N \times N$ matrices with $N>2$, we should extend (\ref{AKNS-Vhier2}) 
to a multicomponent version, see \cite{Dick03}, for example.} 
\vskip.2cm

Equation (\ref{AKNS-Vhier2}) indeed defines a hierarchy, i.e. the 
flows mutually commute, if $\hat{\A}$ admits a direct sum decomposition 
$\hat{\A} = \hat{\A}_{-} \oplus \hat{\A}_{+}$, where 
$\hat{\A}_{-} := (\hat{\A})_{<k}$ and $\hat{\A}_{+} := (\hat{\A})_{\geq k}$ 
are subalgebras, i.e., 
\be
    \hat{\A}_{-} \bullet \hat{\A}_{-} \subset \hat{\A}_{-} \, , \qquad
    \hat{\A}_{+} \bullet \hat{\A}_{+} \subset \hat{\A}_{+}  \; .
       \label{subalg-cond}
\ee
This is the case if $k \in \{ 0,1 \}$. 
Let us now allow deformed associative products, with the indeterminate $\zeta$ as the 
deformation parameter. 
\vskip.2cm 

\noindent
$\mathbf{k=0}$. Let $\bullet_{(0)}$ and $\bullet_{(-1)}$ be two compatible 
associative products in $\A$ (\cite{CGM00,Odes+Soko05,Odes+Soko06}, see 
also appendix~A). Then the new product
\be
    A \bullet B := A \bullet_{(0)} B + A \bullet_{(-1)} B \, \zeta 
\ee
is associative for all $\zeta$ and, extended to $\hat{\A}$, satisfies
\be
    (X_{\geq 0} \bullet Y_{\geq 0})_{\geq 0}
  = X_{\geq 0} \bullet Y_{\geq 0} \, , \qquad
    (X_{<0} \bullet Y_{<0})_{<0} = X_{<0} \bullet Y_{<0} 
\ee
for all $X,Y \in \hat{\A}$, so that the subalgebra conditions (\ref{subalg-cond}) are preserved. Any other $\zeta$-dependence of the 
new product but the linear one would spoil one of these relations. 
Concerning our notation, assigning the index $-1$ to the second product is in 
accordance with an effective $\zeta$-grading.\footnote{For example, taking this 
convention into account, the sum of all indices of each summand of (\ref{k=0:u_nx}) 
or (\ref{k=0:u_1tn}) equals $n+1$. This helps to keep account of the possible terms.} 
\vskip.2cm 

\noindent
$\mathbf{k=1}$. Given two compatibel associative products $\bullet_{(0)}$ and $\bullet_{(1)}$ in $\A$, the new product
\be
    A \bullet B := A \bullet_{(0)} B + A \bullet_{(1)} B \, \zeta^{-1} 
           \label{prod_k=1}
\ee
is associative for all $\zeta$ and, extended to $\hat{\A}$, satisfies
\be
    (X_{\geq 1} \bullet Y_{\geq 1})_{\geq 1}
  = X_{\geq 1} \bullet Y_{\geq 1} \, , \qquad
    (X_{<1} \bullet Y_{<1})_{<1} = X_{<1} \bullet Y_{<1}
\ee
for all $X,Y \in \hat{\A}$, so that (\ref{subalg-cond}) is preserved. 
In fact, the $\zeta$-dependence in (\ref{prod_k=1}) is the only one 
with this property.
\vskip.2cm

In the following we assume that $V$ is given by a dressing
\be
    V = W \bullet P \, \bullet W^{\bullet -1}
    \label{AKNS-dress}
\ee
where $W$ is an invertible formal power series in $\zeta^{-1}$ with 
inverse $W^{\bullet -1}$ (with respect to the product $\bullet$), and
$P \in \A$ is constant and idempotent, i.e.,
\be
     P \bullet P = P \; .        \label{P^bull2}
\ee
Then (\ref{AKNS-dress}) leads to
\be
       V \bullet V = V  \; .    \label{AKNS-Vconstr}
\ee
Introducing $L := V \, \zeta$, this implies\footnote{Here $L^{\bullet n}$ denotes the
$n$-fold product $L \bullet L \cdots \bullet L$. } 
$L^{\bullet n} = \zeta^n \, V$, 
so that the AKNS-type hierarchy (\ref{AKNS-Vhier2}) can be equivalently 
expressed as 
\be
     L_{t_n} = [ (L^{\bullet n})_{\geq k} , L ]_{\bullet}
                \qquad \quad  n=1,2, \ldots  \; .   \label{GD-hier}
\ee
This observation will be important in section~\ref{section:functional}. 
\vskip.2cm

\noindent
{\it Remark.} In view of the structure of (\ref{AKNS-Vhier2}), it seems to be 
more natural and more general to replace the associative algebra $\A$ by a 
Lie algebra which splits into a direct sum of Lie subalgebras. 
Of course, the commutator with respect to the product $\bullet$ in $\A$ induces 
a Lie algebra structure, but not every Lie algebra can be obtained in this way. 
Deformations of Lie algebras in the context of integrable systems have been studied 
in particular in \cite{Golu+Soko02,Golu+Soko04,Golu+Soko05}
and \cite{Skry02,Skry04a,Skry04b,Skry05}. The two classes of Lie algebras 
obtained from the deformations of associative algebras considered above 
are \emph{quasigraded} of type $(1,0)$, respectively $(0,1)$, in the sense of the 
latter references. 
Although in (\ref{P^bull2}) and (\ref{AKNS-Vconstr}) we do make explicit use of an  \emph{associative algebra structure}, the subsequent results in this section 
can also be obtained in a Lie algebraic framework, though with some more 
efforts.\footnote{The methods developed in \cite{Skry02,Skry04a,Skry04b,Skry05} 
can certainly also be used to derive the results of this section.}  
The deeper reason for our choice of the associative algebra framework 
is the applicability of the methods developed in \cite{DMH06func} to compute 
functional representations (see  section~\ref{section:functional}). This choice 
is already required for the conversion of (\ref{AKNS-Vhier2}) into (\ref{GD-hier}). 
\hfill $\blacksquare$

\subsection{The case $k=0$}
Equation (\ref{AKNS-Vconstr}) leads to
\be
    u_0 \bullet_{(-1)} u_0 &=& 0 \, , \label{k=0:u_0-rel} \\
    \sum_{j=0}^n u_j \bullet_{(0)} u_{n-j}
    + \sum_{j=0}^{n+1} u_j \bullet_{(-1)} u_{n-j+1} &=& u_n
      \qquad \quad  n=0,1,2, \ldots  \; .   \label{k=0:u_n-rels}
\ee
The hierarchy equations (\ref{AKNS-Vhier2}) imply, in particular,
\be
    u_{0,x} &=& [ u_0 , u_2 ]_{(-1)}  \, ,  \label{k=0:u_0x}  \\
    u_{n,x} &=& [u_0 , u_{n+1}]_{(0)} + [u_1 , u_n]_{(0)}
                + [ u_0 , u_{n+2} ]_{(-1)} + [u_1 , u_{n+1}]_{(-1)}
     \qquad      n=1,2,\ldots  \qquad     \label{k=0:u_nx}
\ee
and
\be
    u_{0,t_n} &=& [u_0 , u_{n+1}]_{(-1)}  \, , \label{k=0:u_0tn}  \\
    u_{1,t_n} &=& [u_0 , u_{n+1}]_{(0)} + [ u_0 , u_{n+2} ]_{(-1)}
              + [u_1 , u_{n+1}]_{(-1)}  \qquad n=1,2,\ldots \; .
             \label{k=0:u_1tn}
\ee
Combining (\ref{k=0:u_nx}) and (\ref{k=0:u_1tn}), leads to
\be
   u_{1,t_n} =  u_{n,x} - [u_1 , u_n]_{(0)}  \qquad \quad   n=1,2,\ldots \; .
             \label{k=0:u_1-evolution}
\ee
\vskip.2cm

If we assume in addition that
\be
      u_0 = P             \label{u_0=P}
\ee
and
\be
    P \bullet_{(-1)} A = 0 = A \bullet_{(-1)} P   \qquad \forall A \in \A \, ,
        \label{P-assumpt}
\ee
then (\ref{k=0:u_0-rel}), (\ref{k=0:u_n-rels}) for $n=0$,
(\ref{k=0:u_0x}) and (\ref{k=0:u_0tn}) are satisfied (by use of (\ref{P^bull2})).

\subsubsection{Chen-Lee-Liu DNLS hierarchy}
\label{sssection:CLL}
Let $\A$ be a matrix algebra. We define the product $\bullet$ by\footnote{The 
corresponding Lie algebra version is a special case of the Lie algebra 
deformations considered in \cite{Skry04a,Skry05}, for example.}
\be
    A \bullet B := A P B + A (I - P) B \, \zeta
\ee
with the unit matrix $I$ and a fixed matrix $P$. 
For $\zeta = 1$, this is the usual matrix product. It is easily checked
that the deformed product is indeed associative (see also appendix~A).
If $P^2 = P$, then (\ref{P^bull2}) holds and $(\A,\bullet)$ is unital with unit
\be
     J = P + (I - P) \, \zeta^{-1}   \; .
\ee
Furthermore, (\ref{P-assumpt}) holds.
Now we turn to a more concrete example by choosing 
\be
    u_0 = P = \left(\begin{array}{cc} 1 & 0 \\ 0 & 0 \end{array} \right) \; .
\ee
Equation (\ref{k=0:u_n-rels}) with $n=1$ is then solved by
\be
    u_1 = \left(\begin{array}{cc} - q \, r & q \\ r & 0 \end{array} \right) \, , 
         \label{CLL-u_1}
\ee
where $q$ and $r$ are $M \times N$, respectively $N \times M$ matrices,
with entries from a possibly \emph{non}commutative unital algebra 
(over $\mathbb{K}$). 
Equation (\ref{k=0:u_n-rels}) with $n=2$ leads to
\be
    u_2 = \left(\begin{array}{cc} - (q r)^2 - q c - b r & b
            \\ c  & r q \end{array} \right)
\ee
where the entries $b,c$ still have to be determined. Now we use
(\ref{k=0:u_nx}) with $n=1$ to obtain
\be
    u_2 = \left(\begin{array}{cc} q r_x - q_x r + (q r)^2  & q_x - q r q
            \\ -r_x -r q r & r q \end{array} \right) \; .
\ee
Equation (\ref{k=0:u_1-evolution}) for $n=2$ then takes the form\footnote{These 
equations are of `Schr\"odinger type' after replacing $t_2$ by $\imath \, t_2$ with 
the imaginary unit $\imath$. In the case $M=N$, the reductions $q=1$ or $r=1$, 
where $1$ stands for a unit element, lead to noncommutative versions of the Burgers equation.} 
\be
    q_{t_2} - q_{xx} + 2 \, q_x \, r \, q = 0 \, , \qquad
    r_{t_2} + r_{xx} + 2 \, r \, q  \, r_x = 0 \; . 
\ee
These are `noncommutative' (e.g., matrix) versions of a system of coupled
derivative nonlinear Schr\"odinger (DNLS) equations, generalizing the
\emph{Chen-Lee-Liu equation}  \cite{CLL79,KSS95,Tsuc+Wada99}.\footnote{This 
includes the case of vector DNLS equations ($M=1$ or $N=1$). See also 
\cite{APT04} and the references cited therein for the physical relevance of 
vector NLS equations.}
In the next step we obtain
\be
    u_3 = \left(\begin{array}{cc} a  & b
            \\ c & d \end{array} \right)
\ee
where
\be
    a &:=& -q r_{xx} - q_{xx} r + q_x r_x + 2 q_x r q r - 2 q r q r_x
           + q \, d \, r       \, , \\
    b &:=& q_{xx} - 2 q_x r q - q \, d        \, , \\
    c &:=& r_{xx} + 2 r q r_x - d \, r                    \, , \\
    d &:=& r q_x - r_x q - (r q)^2    \; .
\ee
Equation (\ref{k=0:u_1-evolution}) for $n=3$ then becomes the
system
\be
    q_{t_3} - q_{xxx} + 3 \, q_{xx} \, r \, q
         + 3 \, q_x \, r \, q_x - 3 \, q_x \, (r \, q)^2 &=& 0 \, , \\
    r_{t_3} - r_{xxx} - 3 \, r \, q \, r_{xx}
         - 3 \, r_x \, q \, r_x - 3 \, (r \, q)^2 \, r_x &=& 0 \; .
\ee

\subsubsection{Gerdjikov-Ivanov DNLS hierarchy}
\label{sssection:GI}
Let $\A$ be a matrix algebra, $\sigma \in \A$ such that 
\be
    \sigma^2 = I \, ,
\ee
and
\be
    A^{(-)} := \frac{1}{2} (A - \sigma \, A \, \sigma)
      \, , \qquad
    A^{(+)} := A - A^{(-)} \, ,
\ee
for $A \in \A$. This provides us with a decomposition
$\A = \A^{(+)} \oplus \A^{(-)}$.
Then
\be
    A \bullet B := (A B - A^{(-)} B^{(-)}) + A^{(-)} B^{(-)} \, \zeta
\ee
defines an associative deformation of the matrix product (see also 
appendix~A).\footnote{For $2 \times 2$ matrices, the products 
$A B - A^{(-)} B^{(-)}$ and $A^{(-)} B^{(-)}$ appeared as examples 
in \cite{CGM00}, in a different context.}
The unit matrix $I$ is also a unit element with respect to the
deformed product, hence $J = I$. Furthermore, 
$P := (I + \sigma)/2$ satisfies $P^{(-)} = 0$ and thus $P \bullet P = P^2 = P$. 
\vskip.2cm

Choosing
\be
   u_0 = P = \left( \begin{array}{cc} 1 & 0 \\ 0 & 0 \end{array} \right) \, ,
   \qquad
   \sigma = \left( \begin{array}{cc} 1 & 0 \\ 0 & -1 \end{array} \right) \, ,
\ee
(\ref{k=0:u_n-rels}) for $n=1$ leads to
\be
    u_1 = \left(\begin{array}{cc} - q \, r & q \\ r & r \, q \end{array} \right) \; .
          \label{GI-u_1}
\ee
 From (\ref{k=0:u_n-rels}) with $n=2$, and (\ref{k=0:u_nx}), we find
\be
    u_2 = \left(\begin{array}{cc} q \, r_x - q_x \, r - (q \, r)^2 & q_x \\
          - r_x & r \, q_x - r_x \ q + (r \, q)^2 \end{array} \right) \, ,
\ee
According to (\ref{k=0:u_1-evolution}), the first system of
the hierarchy is
\be
    q_{t_2} - q_{xx} - 2 \, q \, r_x \, q + 2 \, (q \, r)^2 \, q &=& 0 \, ,
         \label{GI-q_t2}  \\
    r_{t_2} + r_{xx} - 2 \, r \, q_x \, r - 2 \, r \, (q \, r)^2 &=& 0 \, ,
         \label{GI-r_t2}
\ee
which generalizes the \emph{Gerdjikov-Ivanov equation}, a derivative nonlinear
Schr\"odinger equation with an additional potential term  
\cite{Gerd+Ivan83,Kund87,Olve+Soko98,Fan00,Fan00b,Fan01,GDY01,Tsuc02,Doct02}.

\subsection{The case $k=1$}
 From (\ref{AKNS-Vconstr}), we obtain
\be
    u_0 \bullet_{(0)} u_0 = u_0  \, ,  \label{KN-u_0^2}   \\
    \sum_{j=0}^n u_j \bullet_{(0)} u_{n-j}
    + \sum_{j=0}^{n-1} u_j \bullet_{(1)} u_{n-j-1}
   &=& u_n   \qquad    n=1,2,\ldots \, ,     \label{KN-u_n}
\ee
and the hierarchy equations (\ref{AKNS-Vhier2}) lead to
\be
    u_{n,x} = [u_0,u_{n+1}]_{(0)} + [u_0,u_n]_{(1)} \qquad\quad n=0,1,2,\ldots
    \label{KN-u_nx}
\ee
and
\be
    u_{0,t_n} = [u_0 , u_n]_{(0)} \qquad \quad n=1,2,\ldots  \; .
                 \label{KN-u_0t_n}
\ee
Combined with (\ref{KN-u_nx}), this yields
\be
      u_{0,t_{n+1}} - u_{n,x} + [u_0, u_n]_{(1)} = 0
         \qquad\quad n=0,1,2,\ldots   \; .
\ee

\subsubsection{Kaup-Newell hierarchy}
\label{sssection:KN}
Now we choose the product $\bullet$ as in section~\ref{sssection:GI}, but
with $\zeta$ replaced by $\zeta^{-1}$.\footnote{The choice of the product
used in section~\ref{sssection:CLL} reproduces the results obtained there.}
Hence
\be
    A \bullet_{(0)} B &:=& A B - A^{(-)} B^{(-)}
                     = A^{(+)} B^{(+)} +A^{(+)} B^{(-)} + A^{(-)} B^{(+)}
                    \, , \\
    A \bullet_{(1)} B &:=& A^{(-)} B^{(-)} \; .
\ee
Furthermore, (\ref{KN-u_nx}) and (\ref{KN-u_0t_n}) split as follows,
\be
    u_{n,x}^{(+)} &=& [ u_0^{(+)} , u_{n+1}^{(+)} ] + [ u_0^{(-)} , u_n^{(-)} ] \, ,
           \label{KN-eq1}  \\
    u_{n,x}^{(-)} &=& [ u_0^{(+)} , u_{n+1}^{(-)} ] + [u_0^{(-)} , u_{n+1}^{(+)} ] \, ,
           \label{KN-eq2}
\ee
where $n=0,1,\ldots$, and
\be
    u_{0,t_n}^{(+)} &=& [ u_0^{(+)} , u_n^{(+)} ]  \, , \\
    u_{0,t_n}^{(-)} &=& [ u_0^{(+)} , u_n^{(-)} ] + [u_0^{(-)} , u_n^{(+)} ]
                     = u_{n-1,x}^{(-)} \, ,  \label{KN-eq4}
\ee
where $n=1,2,\ldots$.
Choosing moreover
\be
    P = \left(\begin{array}{cc} 1 & 0 \\ 0 & 0 \end{array}\right) \, ,
    \qquad
    \sigma = \left(\begin{array}{cc} 1 & 0 \\ 0 & -1 \end{array}\right) \, ,
\ee
equation (\ref{KN-u_0^2}) and also (\ref{KN-eq1}) for $n=0$ are solved by
\be
    u_0 = \left(\begin{array}{cc} 1 & q \\ r & 0 \end{array}\right) \; .
          \label{KN-u_0}
\ee
where $q,r$ are ($M \times N$, respectively $N \times M$) matrices with
entries from a, possibly noncommutative, associative algebra.
 From (\ref{KN-u_n}) we obtain $u_1 = u_0 \bullet_{(0)} u_1 + u_1 \bullet_{(0)} u_0
+ u_0 \bullet_{(1)} u_0$. Together with (\ref{KN-eq2}) for $n=0$, this
entails
\be
    u_1 &=& \left(\begin{array}{cc} - q r & q_x - 2 q r q\\ -r_x - 2 r q r &  r q 
            \end{array}\right) \; .
\ee
Now (\ref{KN-eq4}) leads to
\be
    q_{t_2} = q_{xx} - 2 \, (q r q)_x \, , \qquad
    r_{t_2} = - r_{xx} - 2 \, (r q r)_x    \label{KN-system}
\ee
which generalizes the \emph{Kaup-Newell} derivative NLS equation
\cite{Kaup+Newe78,Steu03,Zeng94,Xia+Fan05}.

\section{Functional representations of the derivative NLS hierarchies}
\label{section:functional}
\setcounter{equation}{0}
After recalling a result from \cite{DMH06func}, we apply it to compute 
functional representations of the (matrix) DNLS hierarchies considered 
in section~\ref{section:DNLS&AKNS}.

\subsection{Preliminaries}
\label{ssection:prel}
In \cite{DMH06func} we showed that an integrable hierarchy, from a class which includes 
(\ref{GD-hier}), implies (and is typically equivalent to)
\be
    E(\la_2)_{-[\la_1]} \bullet E(\la_1) = E(\la_1)_{-[\la_2]} \bullet E(\la_2) \, ,
           \label{EE}
\ee
where $\la_1$ and $\la_2$ are indeterminates and
\be
    E(\la) = J + \sum_{n \geq 1} \la^n \, E_n
\ee
is a formal power series with coefficients
\be
    E_n = (\tilde{E}_n)_{+}    \qquad \quad  n=1,2, \ldots \, ,
        \label{E_n}
\ee
recursively determined via
\be
    \tilde{E}_1 = - L \, , \quad
    \tilde{E}_{n+1} = (\tilde{E}_n)_{-} \bullet L
    \qquad \quad  n=1,2, \ldots \; .       \label{tE-recurs}
\ee
Furthermore, we used the following notation (see also \cite{Sato+Sato82,DJM82,Bogd99,DMH06func}, 
for example). For $F$ dependent on $\mathbf{t} := (t_1,t_2,t_3,\ldots)$, we define
\be
    F_{[\la]}(\mathbf{t}) := F(\mathbf{t}+[\la])
       = \sum_{n \geq 0} \la^n \, \chi_n(F) \, , \qquad
    F_{-[\la]}(\mathbf{t}) := F(\mathbf{t}-[\la])
\ee
(as formal power series in $\la$) with
\be
      [\la] := (\la,\la^2/2,\la^3/3,\ldots)
\ee
and
\be
    \chi_n := \mathbf{p}_n(\tpa)  \, , \qquad
    \tpa := (\pa_{t_1}, \pa_{t_2}/2, \pa_{t_3}/3, \ldots)  \, ,
\ee
where $\mathbf{p}_n$, $n=0,1,2,\ldots$, are the elementary Schur polynomials
(see \cite{OSTT88,Macd95,Bogd99}, for example).
\vskip.2cm

It is helpful to express (\ref{EE}) in terms of
\be
    \hat{E}(\la) := E(\la)_{[\la]}  \, ,    \label{hE}
\ee
so that it takes the form
\be
    \hat{E}(\la_2) \bullet \hat{E}(\la_1)_{[\la_2]}
  = \hat{E}(\la_1) \bullet \hat{E}(\la_2)_{[\la_1]} \; .
    \label{hEhE}
\ee
\vskip.2cm

\noindent
{\it Remark.} (\ref{EE}) (or (\ref{hEhE})) should be regarded as a \emph{finite} 
(i.e. group) analogue of an \emph{infinitesimal} (i.e. Lie algebraic) 
zero curvature equation. $E(\la)$ (or $\hat{E}(\la)$) lies in the group of 
invertible elements of the algebra $\hat{\A}$. In certain cases it should be 
possible to regard this group as a (formal) Lie group of a (loop) Lie algebra 
$\hat{\cal G} = \hat{\cal G}_+ \oplus \hat{\cal G}_-$. More precisely, 
$E(\la)$ would then be an element of the group generated by $\hat{\cal G}_+$. 
However, such a Lie algebraic relation may impose unnecessary and inconvenient restrictions on $E(\la)$. In the subsequent calculations it is not at all required. 
\hfill $\blacksquare$

\subsection{The case $k=0$}
We have
\be
   E(\la) = J - (u_0 \, \zeta + u_1) \, \la
            + \sum_{n \geq 2} (\tE_n)_{\geq 0} \, \la^n
\ee
and $E_n = (\tE_n)_{\geq 0}$ is \emph{linear} in $\zeta$ for $n=2,3,\ldots$.
This follows from
\be
    (\tE_{n+1})_{\geq 0}
 &=& ((\tE_n)_{<0} \bullet (u_0 \, \zeta + u_1))_{\geq 0}  \nonumber \\
 &=& \res(\tE_n) \bullet_{(-1)} u_0 \, \zeta
     + \res(\tE_n) \bullet_{(0)} u_0 + \res(\tE_n \zeta) \bullet_{(-1)} u_0
     + \res(\tE_n) \bullet_{(-1)} u_1 \, ,  \qquad    \label{k=0:E_n}
\ee
where $\res(X)$ takes the coefficient of the $\zeta^{-1}$ part of $X$. 
Let us write
\be
    \hat{E}(\la) = J - \Big( v(\la) \, \zeta + w(\la) \Big) \, \la
\ee
where $v(\la) = \sum_{n \geq 0} v_n \, \la^n$ and
$w(\la) = \sum_{n \geq 0} w_n \, \la^n$ are (formal) power series in $\la$
with $v_0 = u_0$ and $w_0 = u_1$.
Now (\ref{hEhE}) splits into the four equations
\be
  &&   v(\la_2) \bullet_{(-1)} v(\la_1)_{[\la_2]}
     = v(\la_1) \bullet_{(-1)} v(\la_2)_{[\la_1]} \, ,  \\
  &&   v(\la_2) \bullet_{(0)} v(\la_1)_{[\la_2]}
     + v(\la_2) \bullet_{(-1)} w(\la_1)_{[\la_2]}
     + w(\la_2) \bullet_{(-1)} v(\la_1)_{[\la_2]}   \nonumber \\
 &=& v(\la_1) \bullet_{(0)} v(\la_2)_{[\la_1]}
     + v(\la_1) \bullet_{(-1)} w(\la_2)_{[\la_1]}
     + w(\la_1) \bullet_{(-1)} v(\la_2)_{[\la_1]} \, ,   \\
  && \la_1^{-1} \left( v(\la_2)_{[\la_1]} - v(\la_2) \right)
     - \la_2^{-1} \left( v(\la_1)_{[\la_2]} - v(\la_1) \right)   \nonumber \\
 &=& v(\la_1) \bullet_{(0)} w(\la_2)_{[\la_1]}
     - v(\la_2) \bullet_{(0)} w(\la_1)_{[\la_2]}
     + w(\la_1) \bullet_{(0)} v(\la_2)_{[\la_1]}
     - w(\la_2) \bullet_{(0)} v(\la_1)_{[\la_2]} \nonumber \\
  && + w(\la_1) \bullet_{(-1)} w(\la_2)_{[\la_1]}
     - w(\la_2) \bullet_{(-1)} w(\la_1)_{[\la_2]} \, ,   \\
  && \la_1^{-1} \left( w(\la_2)_{[\la_1]} - w(\la_2) \right)
     - \la_2^{-1} \left(w(\la_1)_{[\la_2]} - w(\la_1) \right)  \nonumber \\
  &=& w(\la_1) \bullet_{(0)} w(\la_2)_{[\la_1]}
     - w(\la_2) \bullet_{(0)} w(\la_1)_{[\la_2]} \; .
            \label{k=0:f1}
\ee
Assuming (\ref{u_0=P}) and (\ref{P-assumpt}), equation (\ref{k=0:E_n}) shows that
\be
     v(\la) = P \, ,
\ee
and the first two equations are identically satisfied. Furthermore, the third
equation reduces to
\be
 & &  P \bullet_{(0)} \Big( w(\la_2)_{[\la_1]} - w(\la_1)_{[\la_2]} \Big)
       + \Big( w(\la_1) - w(\la_2) \Big) \bullet_{(0)} P    \nonumber  \\
 &=& w(\la_2) \bullet_{(-1)} w(\la_1)_{[\la_2]}
      - w(\la_1) \bullet_{(-1)} w(\la_2)_{[\la_1]} \; .  \label{k=0:f2}
\ee
In the limit $\la_2 \to 0$, the equations (\ref{k=0:f1}) and (\ref{k=0:f2})
lead to
\be
    \la^{-1} (w_{0,[\la]}-w_0) - w(\la)_x
  = w(\la) \bullet_{(0)} w_{0,[\la]} - w_0 \bullet_{(0)} w(\la) \, ,
    \label{k=0:f1lim}
\ee
respectively
\be
    P \bullet_{(0)} (w_{0,[\la]} - w(\la)) + (w(\la)-w_0) \bullet_{(0)} P
  = w_0 \bullet_{(-1)} w(\la) - w(\la) \bullet_{(-1)} w_{0,[\la]} \; .
    \label{k=0:f2lim}
\ee

\subsubsection{Functional representation of the Chen-Lee-Liu DNLS hierarchy}
\label{sssection:CLL2}
We choose $\A$, the product $\bullet$, and $P$ as in section~\ref{sssection:CLL}.
Let us now turn to the derivation of a corresponding functional representation
of the hierarchy. We write
\be
    w(\la) = \left(\begin{array}{cc} p(\la) & q(\la) \\ r(\la) & s(\la)
             \end{array}\right),    \label{DAKNS_N=2_w}
\ee
where the entries are power series in $\la$.
In this case (\ref{k=0:E_n}) reads
\be
    E_{n+1} = (\tE_{n+1})_{\geq0} = \res(\tE_n) P + \res(\tE_n) (I-P) u_1 \; .
\ee
Since $(u_1)_{22} = 0$ according to (\ref{CLL-u_1}), we easily deduce from this formula that
\be
    s(\la) = 0 \; .
\ee
Now (\ref{k=0:f1}) and (\ref{k=0:f2}) are turned into the following
equations,
\be
       \la_2^{-1} (p(\la_1)_{[\la_2]} - p(\la_1))
     - \la_1^{-1} (p(\la_2)_{[\la_1]} - p(\la_2))
 &=& p(\la_2) \, p(\la_1)_{[\la_2]} - p(\la_1) \, p(\la_2)_{[\la_1]} \, ,
                                 \label{CLL-pla} \\
       \la_2^{-1} (q(\la_1)_{[\la_2]} - q(\la_1))
     - \la_1^{-1} (q(\la_2)_{[\la_1]} - q(\la_2))
 &=& p(\la_2) \, q(\la_1)_{[\la_2]} - p(\la_1) \, q(\la_2)_{[\la_1]} \, ,
                                 \label{CLL-qla} \\
       \la_2^{-1} (r(\la_1)_{[\la_2]} - r(\la_1))
     - \la_1^{-1} (r(\la_2)_{[\la_1]} - r(\la_2))
 &=& r(\la_2) \, p(\la_1)_{[\la_2]} - r(\la_1) \, p(\la_2)_{[\la_1]} \, ,
                                 \label{CLL-rla} \\
 0 &=& r(\la_1) \, q(\la_2)_{[\la_1]} - r(\la_2) \, q(\la_1)_{[\la_2]} \, ,
                                 \label{CLL-sla}
\ee
and
\be
      p(\la_1)_{[\la_2]} - p(\la_1) - p(\la_2)_{[\la_1]} + p(\la_2)
  &=& q(\la_1) \, r(\la_2)_{[\la_1]} - q(\la_2) \, r(\la_1)_{[\la_2]} \, ,
              \label{CLL-plalim} \\
      q(\la_1)_{[\la_2]} - q(\la_2)_{[\la_1]}
  &=& 0 \, ,     \label{fq} \\
      r(\la_2) - r(\la_1) &=& 0 \; .
              \label{CLL-rlalim}
\ee
As a consequence of (\ref{fq}) and (\ref{CLL-rlalim}), we have
\be
    q(\la) = q_{[\la]} \, , \qquad
    r(\la) = r  \, ,   \label{CLL-qla,rla}
\ee
and (\ref{CLL-sla}) is automatically satisfied. Now (\ref{CLL-plalim}) becomes
\be
    p(\la_1)_{[\la_2]} - p(\la_1) - p(\la_2)_{[\la_1]} + p(\la_2)
  = (q r)_{[\la_1]} - (q r)_{[\la_2]} \; .
\ee
In the limit $\la_2 \to 0$, this yields $-p_{0[\la]}+p_0 = (qr)_{[\la]}-qr$,
which is solved by $p_0 = -q r$, in accordance with the upper left entry of
$u_1$ in (\ref{CLL-u_1}).
The coefficient of the $\la_1^n \la_2^m$ term, $m,n \in \mathbb{N}$, of the
above equation reads $\chi_n(p_m) = \chi_m(p_n)$. Hence there is a $p$ such
that
\be
     p_n = \chi_n(p) \qquad \quad   n = 1,2,\ldots \, ,
               \label{CLL-p_n=chi_n(p)}
\ee
and consequently
\be
    p(\la) = p_{[\la]} - p - q \, r \; .    \label{CLL-p(la)}
\ee
\vskip.2cm

Taking the limit $\la_2 \to 0$ of (\ref{CLL-pla}), (\ref{CLL-qla}) and (\ref{CLL-rla}),
and using (\ref{CLL-qla,rla}), we obtain
\be
     p(\la)_x &=& - \la^{-1} \left( (q r)_{[\la]} - q \, r \right)
                   + p(\la) \, (q \, r)_{[\la]} - q \, r \, p(\la) \, ,
                          \label{CLL-pla_x} \\
  q_{[\la],x} &=& \la^{-1} (q_{[\la]} - q )
                - \left( q \, r + p(\la) \right) \, q_{[\la]} \, ,
                          \label{CLL-q_la_x} \\
          r_x &=& \la^{-1} (r_{[\la]} - r)
                  + r \, \left( p(\la) + (q \, r)_{[\la]} \right) \; .
                          \label{CLL-r_x}
\ee
Multiplying (\ref{CLL-q_la_x}) by $r$ from the right and
(\ref{CLL-r_x}) by $q_{[\la]}$ from the left, adding the results, and
using (\ref{CLL-p(la)}), we get
\be
    (q_{[\la]} \, r)_x
  = (\la^{-1} + q_{[\la]} \, r) \left( (q r)_{[\la]} - q \, r \right)
    + [q_{[\la]} \, r, p_{[\la]} - p] \; .
\ee
With the help of (\ref{CLL-p(la)}) we rewrite (\ref{CLL-pla_x}) as follows,
\be
    p(\la)_x = - (\la^{-1} - p(\la)) \left( (q \, r)_{[\la]} - q \, r \right)
               + [ p(\la) , p_{[\la]}-p ] \; .
\ee
Adding the last two equations, provides us with
\be
    (p(\la) + q_{[\la]} \, r)_x
  = (p(\la) + q_{[\la]} \, r) \left( (q \, r)_{[\la]} - q \, r \right)
            + [ p(\la) + q_{[\la]} \, r , p_{[\la]} - p ] \; .
\ee
Expanding in powers of $\la$, using $p_0 = - q \, r$, and setting possible
integration constants to zero\footnote{This would be enforced by imposing an  
auxiliary boundary condition like $w(\la) \to 0$ as $x \to \pm \infty$.}, 
this leads to
\be
    p(\la) = - q_{[\la]} \, r  \; .
\ee
Inserting this in (\ref{CLL-q_la_x}) and (\ref{CLL-r_x}),
we obtain the following functional representation of the hierarchy,
\be
    (q - q_{-[\la]}) (\la^{-1} + r_{-[\la]} \, q) &=& q_x \, ,  \\
    (\la^{-1} + r \, q_{[\la]}) (r_{[\la]} - r) &=& r_x \; .
\ee
By differentiation with respect to $\la$, we recover the corresponding
equations in \cite{Veks02}, obtained in the case of commuting variables.
\vskip.2cm

\noindent
{\it Remark.}
Expanding (\ref{CLL-pla}) and using (\ref{CLL-p_n=chi_n(p)}),
the coefficients of $\la_1^m \la_2^n$, $m,n \geq 1$, entail
\be
   \chi_{n+1}(\chi_m(p)) - \chi_{m+1}(\chi_n(p))
 = \sum_{j=1}^n \chi_j(p) \, \chi_{n-j}(\chi_m(p))
     - \sum_{j=1}^m \chi_j(p) \, \chi_{m-j}(\chi_n(p)) \; .
\ee
This system is equivalent to the potential KP hierarchy (see section~3.3 in
\cite{DMH06func}). Since (\ref{CLL-p(la)}) implies $p_x = - q \, r_x$,
it follows that
\be
    p = - \int q_x \, r \, dx
\ee
solves the potential KP hierarchy if $q$ and $r$ satisfy the above
Chen-Lee-Liu DNLS hierarchy.
\hfill $\blacksquare$

\subsubsection{Functional representation of the Gerdjikov-Ivanov DNLS hierarchy}
\label{sssection:GI2}
We choose $\A$, the product $\bullet$, and $P$ as in section~\ref{sssection:GI}.
In order to determine a functional representation of the hierarchy,
we write again
\be
    w(\la) = \left(\begin{array}{cc} p(\la) & q(\la) \\
    r(\la) & s(\la) \end{array} \right)
\ee
where the entries are (formal) power series in $\la$, and obtain from
(\ref{k=0:f2}), as in section~\ref{sssection:CLL2},
\be
    q(\la) = q_{[\la]} \, , \qquad
    r(\la) = r \, , \qquad
    p(\la) = p_{[\la]} - p - q \, r \; .    \label{GI-qla,rla,pla}
\ee
 From (\ref{k=0:f1}) we obtain the system
\be
         \la_1^{-1} (p(\la_2)_{[\la_1]} - p(\la_2))
       - \la_2^{-1} (p(\la_1)_{[\la_2]} - p(\la_1))
    &=& p(\la_1) \, p(\la_2)_{[\la_1]} - p(\la_2) \, p(\la_1)_{[\la_2]} \, ,
              \label{GI-gip} \\
         \la_1^{-1} (s(\la_2)_{[\la_1]} - s(\la_2))
       - \la_2^{-1} (s(\la_1)_{[\la_2]} - s(\la_1))
    &=& s(\la_1) \, s(\la_2)_{[\la_1]} - s(\la_2) \, s(\la_1)_{[\la_2]} \, ,
               \label{GI-gis} \\
        \la_1^{-1} (q_{[\la_1]+[\la_2]} - q_{[\la_2]})
         - \la_2^{-1} \left( q_{[\la_1]+[\la_2]} - q_{[\la_1]} \right)
    &=& (p(\la_1) - p(\la_2)) \, q_{[\la_1]+[\la_2]} \nonumber \\
    & & + q_{[\la_1]} \, s(\la_2)_{[\la_1]} - q_{[\la_2]} \, s(\la_1)_{[\la_2]} \, ,
                 \label{GI-giq}
\ee
and
\be
      \la_1^{-1} (r_{[\la_1]} - r) - \la_2^{-1} (r_{[\la_2]} - r)
    = r \, \left( p(\la_2)_{[\la_1]} - p(\la_1)_{[\la_2]} \right)
      + s(\la_1) \, r_{[\la_1]} - s(\la_2) \, r_{[\la_2]} \; . \label{GI-gir}
\ee
The $\la_1 \to 0$ limit of these equations leads to
\be
    \left( \la^{-1} - p(\la) \right)_x
  &=& \left( \la^{-1} - p(\la) \right) \left( (q r)_{[\la]}- q r \right)
    + \left[ \la^{-1} - p(\la) , p_{[\la]} - p \right] \, , \label{GI-dgip}  \\
          s(\la)_x
  &=& \la^{-1} [ (r q)_{[\la]} - r q ] + r \, q \, s(\la)
               - s(\la) \, (r q)_{[\la]} \, ,     \label{GI-dgis}
\ee
and
\be
  q_{[\la],x} &=& \la^{-1} (q _{[\la]} - q )
      - \left( p(\la) + (q \, r)_{[\la]} + q \, r \right) \, q_{[\la]}
                    + q \, s(\la) \, ,  \label{GI-qla_x}  \\
    r_x &=& \la^{-1} (r_{[\la]} - r)
      + r \, \left( p(\la) + (q \, r)_{[\la]} + q \, r \right)
              - s(\la) \, r_{[\la]} \; .   \label{GI-r_x}
\ee
Here we made use of $s_0 = r \, q$, $p_0 = - q \, r$ (see
(\ref{GI-u_1})), and (\ref{GI-qla,rla,pla}).
Multiplying (\ref{GI-qla_x}) by $r$ from the left and (\ref{GI-r_x})
by $q_{[\la]}$ from the right, and adding the results, we get
\be
     (r \, q_{[\la]})_x
  = \la^{-1} [(r \, q)_{[\la]} - r \, q ] + r \, q \, s(\la)
     - s(\la) (r \, q)_{[\la]} \; .
\ee
Comparing this with (\ref{GI-dgis}), we obtain
\be
    s(\la) = r \, q_{[\la]} \, ,
\ee
setting a constant of integration to zero.
Now (\ref{GI-qla_x}) and (\ref{GI-r_x}) take the form
\be
    q_{[\la],x} &=& \la^{-1} (q _{[\la]} - q )
      - \left( p(\la) + (q \, r)_{[\la]} \right) \, q_{[\la]} \, ,
               \label{GI-dgiq} \\
    r_x &=& \la^{-1} (r_{[\la]} - r) + r \, (p(\la) + q \, r) \; .
               \label{GI-dgir}
\ee
Multipling the first equation by $r$ from the right and the second by $q_{[\la]}$
from the left, adding the results, and using the last of the relations (\ref{GI-qla,rla,pla}), leads to
\be
    \left( \la^{-1} -q_{[\la]} \, r \right)_x
  = - \left( (q \, r)_{[\la]} - q \, r \right) \left( \la^{-1} - q_{[\la]} \, r \right)
    + \left[ \la^{-1} - q_{[\la]} \, r , p_{[\la]} - p \right] \; .
\ee
We multiply this in turn by $\la^{-1} - p(\la)$ from the left, and (\ref{GI-dgip})
by $\la^{-1} - q_{[\la]} \, r$ from the right, and add the results to obtain
\be
    \left[ \left( \la^{-1} - p(\la) \right)
      \left( \la^{-1} - q_{[\la]} \, r \right) \right]_x
  = \left[ \left( \la^{-1} - p(\la) \right)
      \left( \la^{-1} - q_{[\la]} \, r \right) , p_{[\la]} - p \right] \; .
\ee
Integrating this order by order in $\la$, leads to
\be
    \left( 1 - \la \, p(\la) \right) \left( 1 - \la \, q_{[\la]} \, r \right)
    = 1 \, ,
\ee
or equivalently
\be
    p(\la) = - q_{[\la]} \, r \, ( 1 - \la \, q_{[\la]} \, r )^{-1}
           = - ( 1 - \la \, q_{[\la]} \, r )^{-1} \, q_{[\la]} \, r  \; .
\ee
Inserting this expression in (\ref{GI-dgiq}) and (\ref{GI-dgir}),
yields the following functional form of the hierarchy,
\be
    \la^{-1} (q - q_{-[\la]})
  - q \, \Big( r - r_{-[\la]} \, (1 - \la \, q \, r_{-[\la]} )^{-1} \Big) \, q
        &=& q_x \, , \\
    \la^{-1} (r_{[\la]} - r)
  + r \, \Big( q - (1 - \la \, q_{[\la]} \, r)^{-1} \, q_{[\la]} \Big) \, r
        &=& r_x \; .
\ee
Again, we should stress that $q$ and $r$ may be $M \times N$ and $N \times M$
matrices of functions (or, more generally, matrices with elements of any
(noncommutative) associative algebra), where $M,N \in \mathbb{N}$.
\vskip.2cm

\noindent
{\it Remark.} In the same way as in section~\ref{sssection:CLL2}
(see the remark there), we find that
\be
       p = - \int \left( q_x \, r + (q \, r)^2 \right) \, \d x
\ee
satisfies the potential KP hierarchy as a consequence of the above
Gerdjikov-Ivanov DNLS hierarchy.
\hfill $\blacksquare$

\subsection{The case $k=1$}
According to (\ref{E_n}) and (\ref{tE-recurs}), we have
\be
   E_0 = J \, , \qquad
   E_1 = - L_{\geq 1} = - u_0 \, \zeta \, , \qquad
   E_2 = ((\tE_1)_{<1} \bullet L)_{\geq 1} = - u_1 \bullet_{(0)} u_0 \, \zeta \, ,
\ee
and
\be
    E_{n+1} = ( (\tE_n)_{<1} \bullet L)_{\geq 1}
            = ( (\tE_n)_{<1} \bullet u_0 \, \zeta)_{\geq 1}
            = \res(\tE_n \zeta^{-1}) \bullet_{(0)} u_0  \, \zeta
            \qquad n=1,2, \ldots \; .
\ee
Hence
\be
    \hat{E}(\la) = E(\la)_{[\la]} = J - \la \, w(\la) \, \zeta \, , \qquad
    w(\la) = u_0 + \la u_1 \bullet_{(0)} u_0 + \cdots \, ,
              \label{KN-Ew}
\ee
where $w(\la)$ does not depend on $\zeta$.
Now (\ref{hEhE}) yields the two equations
\be
   w(\la_2) \bullet_{(0)} w(\la_1)_{[\la_2]}
  - w(\la_1) \bullet_{(0)} w(\la_2)_{[\la_1]} = 0 \; .
    \label{KN-f1}
\ee
and
\be
  &&   \la_2^{-1} \left( w(\la_1)_{[\la_2]} - w(\la_1) \right)
     - \la_1^{-1} \left( w(\la_2)_{[\la_1]} - w(\la_2) \right) \nonumber \\
 &=&   w(\la_2) \bullet_{(1)} w(\la_1)_{[\la_2]}
     - w(\la_1) \bullet_{(1)} w(\la_2)_{[\la_1]}     \label{KN-f2}
\ee
In the limit $\la_2 \to 0$, this becomes
\be
    u_0 \bullet_{(0)} w(\la) = w(\la) \bullet_{(0)} u_{0,[\la]}
\ee
and
\be
    w(\la)_x - \la^{-1} (u_{0,[\la]} - u_0)
  = u_0 \bullet_{(1)} w(\la) - w(\la) \bullet_{(1)} u_{0,[\la]} \; .
\ee

\subsubsection{Functional representation of the Kaup-Newell hierarchy}
We choose $\A$, the product $\bullet$, and $P$ as in section~\ref{sssection:KN}.
Let us write
\be
    w(\la) = \left(\begin{array}{cc} p(\la) & q(\la) \\
                      r(\la) & s(\la) \end{array} \right)
\ee
where the entries are formal power series in $\la$.
Since $(u_0)_{22} = 0$ and $E_{n+1} = \res( \tE_n \zeta^{-1}) \bullet_{(0)} u_0$,
we have
\be
    s(\la) = 0 \; .
\ee
Furthermore, $q_0 = q$, $r_0 = r$, and $p_0 = 1$ according to (\ref{KN-u_0}).
 From (\ref{KN-f1}) and (\ref{KN-f2}) we obtain
\be
       p(\la_2) \, p(\la_1)_{[\la_2]} - p(\la_1) \, p(\la_2)_{[\la_1]}
   &=& 0 \, ,  \label{KN-tw2_11} \\
       p(\la_2) \, q(\la_1)_{[\la_2]} - p(\la_1) \, q(\la_2)_{[\la_1]}
   &=& 0 \, ,  \label{KN-tw2_12}  \\
       r(\la_2) \, p(\la_1)_{[\la_2]} - r(\la_1) \, p(\la_2)_{[\la_1]}
   &=& 0  \, ,   \label{KN-tw2_21}
\ee
and
\be
          \la_2^{-1} \left( p(\la_1)_{[\la_2]} - p(\la_1) \right)
       - \la_1^{-1} \left( p(\la_2)_{[\la_1]} - p(\la_2) \right)
   &=& q(\la_2) \, r(\la_1)_{[\la_2]} - q(\la_1) \, r(\la_2)_{[\la_1]} \, ,
                     \label{KN-tw1_11} \\
          \la_2^{-1} \left( q(\la_1)_{[\la_2]} - q(\la_1) \right)
       - \la_1^{-1} \left( q(\la_2)_{[\la_1]} - q(\la_2) \right)
   &=& 0 \, ,    \label{KN-tw1_12} \\
          \la_2^{-1} \left( r(\la_1)_{[\la_2]} - r(\la_1) \right)
       - \la_1^{-1} \left( r(\la_2)_{[\la_1]} - r(\la_2) \right)
   &=& 0 \, , \label{KN-tw1_21} \\
          r(\la_2) \, q(\la_1)_{[\la_2]} - r(\la_1) \, q(\la_2)_{[\la_1]}
   &=& 0 \, ,   \label{KN-tw1_22}
\ee
respectively.
Equation (\ref{KN-tw2_11}) (with $p_0 = 1$) is solved by
\be
    p(\la) = f \, f_{[\la]}^{-1}    \label{KN-pla-f}
\ee
with some invertible $f$. Inserting this in (\ref{KN-tw2_12}), we obtain
\be
     \left( f^{-1} q(\la_2) \right)_{[\la_1]}
   = \left( f^{-1} q(\la_1) \right)_{[\la_2]} \, ,
\ee
which implies
\be
     f^{-1} q(\la) = \tq_{[\la]}    \label{KN-tq_la}
\ee
with some $\tq$ which is determined by taking the limit $\la \to 0$,
\be
    \tq = f^{-1} q \; .
\ee
Inserting (\ref{KN-pla-f}) in (\ref{KN-tw2_21}), we obtain
$r(\la_2) \, f_{[\la_2]} = r(\la_1) \, f_{[\la_1]}$ and thus
\be
    r(\la) =  \tir \, f^{-1}_{[\la]}
    \qquad \mbox{with} \qquad
    \tir := r \, f \; .
    \label{KN-r(la)}
\ee
As a consequence, (\ref{KN-tw1_22}) is automatically satisfied,
and (\ref{KN-tw1_11}) takes the form
\be
    \la_2^{-1} \Big( f^{-1} f_{[\la_2]} - (f^{-1} f_{[\la_2]})_{[\la_1]} \Big)
  - \la_1^{-1} \Big( f^{-1} f_{[\la_1]} - (f^{-1} f_{[\la_1]})_{[\la_2]} \Big)
  = (\tq \tir)_{[\la_2]} - (\tq \tir)_{[\la_1]} \; .  \label{KN-f-qr}
\ee
In the limit $\la_2 \to 0$, this yields
\be
    f^{-1} f_x - (f^{-1} f_x)_{[\la]}  = \tq \, \tir - (\tq \, \tir)_{[\la]} \, ,
\ee
and thus
\be
    f^{-1} f_x = \tq \, \tir \, ,  \label{KN-f_x}
\ee
up to addition of a constant, which we set to zero.
The last equation is equivalent to $p_1 = - q \, r$.
Expansion of (\ref{KN-f-qr}) leads to
\be
   \chi_n(f^{-1} \chi_{m+1}(f)) = \chi_m(f^{-1} \chi_{n+1}(f))
     \qquad \quad    m,n = 1,2,\ldots \, ,
\ee
which states that $f$ has to solve the mKP hierarchy. 
Indeed, the last equation is a formulation of the mKP hierarchy (see \cite{DMH06func}). 
It implies the existence of a potential $\phi$ such that
\be
    \chi_{n+1}(f) = - f \, \chi_n(\phi)  \qquad \quad   n=1,2,\ldots \, ,
\ee
which, in functional form, reads
\be
    f_{[\la]} = f + \la \, f_x - \la \, f \, (\phi_{[\la]} - \phi) \; .
\ee
Hence
\be
  f^{-1} f_{[\la]} = 1 + \la \, \left( \tq \, \tir - \phi_{[\la]} + \phi \right) \; .
       \label{KN-f-phi-eq1}
\ee
Using (\ref{KN-f_x}) and the last equation, we find
\be
       (f^{-1} f_{[\la]})_x
  &=& - f^{-1} f_x \, f^{-1} f_{[\la]} + f^{-1} f_{[\la]} \, (f^{-1} f_x)_{[\la]}
            \nonumber \\
  &=& (f^{-1} f_{[\la]}) \left( (\tq \tir)_{[\la]} - \tq \tir \right)
      + [ f^{-1} f_{[\la]} , \tq \tir ]  \nonumber \\
  &=& (f^{-1} f_{[\la]}) \left( (\tq \tir)_{[\la]} - \tq \tir \right)
      + [ f^{-1} f_{[\la]} , \phi_{[\la]} - \phi] \; .   \label{KN-f-phi-eq2}
\ee
Furthermore, the $\la_2 \to 0$ limits of (\ref{KN-tw1_12})
and (\ref{KN-tw1_21}) are
\be
    q(\la)_x = \la^{-1} (q_{[\la]} - q) \, , \qquad
    r(\la)_x = \la^{-1} (r_{[\la]} - r)  \; .    \label{KN-qla,rla_x}
\ee
Conversely, these equations imply (\ref{KN-tw1_12}) and (\ref{KN-tw1_21}),
since the $x$-derivatives of the latter are satisfied as a consequence
of (\ref{KN-qla,rla_x}).
Using (\ref{KN-f_x}) and (\ref{KN-f-phi-eq1}), (\ref{KN-qla,rla_x}) becomes
\be
    \tq_{[\la],x} &=& {1 \over \la} (\tq_{[\la]} - \tq)
        - (\phi_{[\la]} - \phi) \, \tq_{[\la]}  \, ,       \\
    \tir_x &=& {1 \over \la} (\tir_{[\la]} - \tir)
        + \tir \, \left( (\tq \tir)_{[\la]} - \tq \, \tir \right)
        + \tir \, ( \phi_{[\la]} - \phi )  \; .
\ee
Multiplying the first equation by $\tir$ from the right, the second by
$\tq_{[\la]}$ from the left, and adding the results, we obtain
\be
    (\la^{-1} + \tq_{[\la]} \, \tir)_x
  = (\la^{-1} + \tq_{[\la]} \, \tir) \left( (\tq \tir)_{[\la]} - \tq \, \tir \right)
     + [ \tq_{[\la]} \, \tir , \phi_{[\la]} - \phi ] \; .
\ee
Now we multiply (\ref{KN-f-phi-eq2}) by $\la^{-1}$ and subtract the resulting
equation from the preceding one, to get
\be
     z(\la)_x
   = z(\la) \, \left( (\tq \tir)_{[\la]} - \tq \, \tir \right)
     + [ z(\la) , \phi_{[\la]} - \phi ]
\ee
with the following power series in $\la$,
\be
    z(\la) := \la^{-1} (1 - f^{-1} f_{[\la]}) + \tq_{[\la]} \, \tir \; .
\ee
Expanding in powers of $\la$, a simple induction argument (using (\ref{KN-f_x}), 
and setting $x$-integration constants to zero) now leads to $z(\la)=0$ and thus
\be
    f^{-1} f_{[\la]} = 1 + \la \, \tq_{[\la]} \, \tir \, ,  \label{KN-pre-p(la)}
\ee
which, with the help of (\ref{KN-tq_la}) and (\ref{KN-r(la)}),
can be written in the form\footnote{In the case of \emph{commuting} variables,
this formula can be derived in a much simpler way.}
\be
    p(\la) = 1 - \la \, q(\la) \, r(\la) \; .
\ee
 From (\ref{KN-tq_la}) and (\ref{KN-r(la)}), using (\ref{KN-pla-f}), we get
\be
    q(\la) = p(\la) \, q_{[\la]} = ( 1 - \la \, q(\la) \, r(\la) ) \, q_{[\la]}
       \, , \qquad
    r(\la) = r \, p(\la) = r \, ( 1 - \la \, q(\la) \, r(\la) ) \; .
       \label{KN-qla,rla}
\ee
Introducing potentials via
\be
        q = Q_x \, , \qquad
        r = R_x \, ,
\ee
the equations (\ref{KN-qla,rla_x}) can be integrated to give
\be
    q(\la) = \la^{-1} (Q_{[\la]} - Q) \, , \qquad
    r(\la) = \la^{-1} (R_{[\la]} - R) \; .
\ee
Inserting this in (\ref{KN-qla,rla}), we arrive at the
following functional representation of the hierarchy,
\be
      (Q - Q_{-[\la]}) \left( 1 + (R - R_{-[\la]}) \, Q_x \right)
  &=& \la \, Q_x \ , \\
      \left( 1 + R_x \, (Q_{[\la]} - Q) \right) (R_{[\la]} - R)
  &=& \la \, R_x \; .
\ee
Expansion in powers of $\la$ leads, in lowest order, to
\be
    Q_{t_2} = Q_{xx} - 2 \, Q_x \, R_x \, Q_x \, , \qquad
    R_{t_2} = - R_{xx} - 2 \, R_x \, Q_x \, R_x \, ,
\ee
which is the potential form of (\ref{KN-system}).

\section{Conclusions}
\label{section:concl}
\setcounter{equation}{0}
Via deformations of associative products, we expressed derivative NLS 
hierarchies in the form of AKNS hierarchies and then also as `Gelfand-Dickey-type' 
hierarchies (in the sense of (\ref{GD-hier})). 
We then applied a method to compute functional representations of 
`Gelfand-Dickey-type' integrable hierarchies to the derivative NLS hierarchies. 
For the Chen-Lee-Liu DNLS hierarchy, 
we recovered corresponding formulae obtained previously by Vekslerchik \cite{Veks02}.
The functional representations obtained in case of the other derivative 
NLS hierarchies did not yet appear in the literature, according to our knowledge.
In any case, the generalizations to noncommuting dependent variables 
appear to be new. 
\vskip.2cm

The results of \cite{DMH06func} and the present work also demonstrate that 
certain properties of integrable hierarchies are surprisingly 
easily obtained from the formulation in terms of $E(\la)$, 
which has to solve the zero curvature conditions in the form (\ref{EE}). 
Regarding $E(\la)$ as a parallel transport operator taking objects at $\mathbf{t}$ 
to objects at $\mathbf{t}-[\la]$, equation (\ref{EE}) attains the interpretation of a 
\emph{`discrete'} zero curvature condition\footnote{Equation (\ref{EE}) can be interpreted 
as `discrete' in the sense of \cite{DJM82}. See also \cite{ABS03,Bobe+Suri02} 
for a general approach towards integrable systems via discrete zero curvature equations.}, 
which is depicted in the following (commutative) diagram.

\begin{figure}[h]
\begin{center}
\includegraphics{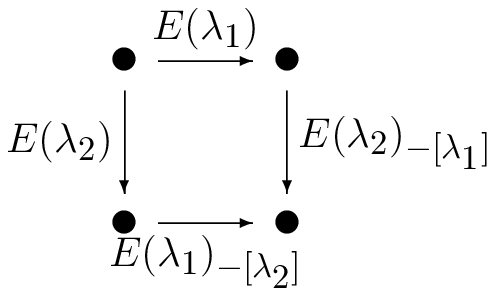}
\end{center}
\end{figure}

\noindent
Equation (\ref{EE}) has the following gauge invariance,
\be
   E(\la) \quad \mapsto \quad G_{-[\la]} \bullet E(\la) \bullet G^{\bullet -1}
\ee
with an invertible $G \in \A$. If one finds such a transformation with 
a nontrivial dependence on $\zeta$, it gives rise to a B\"acklund transformation. 
Corresponding examples and further explorations of (\ref{EE}) and its 
multi-component generalizations will be presented in a separate work.

\renewcommand{\theequation} {\Alph{section}.\arabic{equation}}
\renewcommand{\thelemma} {\Alph{section}.\arabic{lemma}}
\renewcommand{\theproposition} {\Alph{section}.\arabic{proposition}}

\section*{Appendix A: Some elementary notes on deformations of associative algebras}
\addcontentsline{toc}{section}{\numberline{}Appendix A: Notes on
deformations of associative algebras}
\setcounter{section}{1}
\setcounter{equation}{0}
Although in the context of this work the question of \emph{inequivalent}
deformations of associative products addressed by the cohomological methods of 
deformation theory \cite{Gers64,Gers05} appears to be of minor importance, 
corresponding considerations help to find particular deformations  
(see also \cite{CGM00,Odes+Soko05}, and \cite{Golu+Soko02} for the Lie algebra case).  
The following is an elementary approach and we refer the reader to the literature
for more substantial treatments.
Let $\A$ be an associative algebra. Here we are only looking for
associative deformations depending \emph{linearly} on a parameter, say $\zeta$,
\be
    A \bullet B := A \, B + \zeta \, \Phi(A,B) \, ,
\ee
where $\Phi$ is a bilinear mapping $\A \times \A \rightarrow \A$.
Several examples appeared in the main part of this work.
Associativity imposes the following conditions on $\Phi$,
\be
    A \, \Phi(B,C) - \Phi(A B,C) + \Phi(A,B C) - \Phi(A,B) \, C = 0
         \label{Phi-cob}
\ee
and
\be
    \Phi(A,\Phi(B,C)) = \Phi(\Phi(A,B),C) \; .   \label{Phi-assoc}
\ee
The last relation means that $\Phi$ defines an associative product, the other
relation tells us that $\Phi$ is a coboundary. 
This is a \emph{compatibility condition} for the two products. 
If $\Phi$ is a cocycle, so that there is a linear mapping 
$N : \A \rightarrow \A$ such that
\be
    \Phi(A,B) = N(A) \, B + A \, N(B) - N(A B)   \, ,
\ee
then (\ref{Phi-cob}) is solved and (\ref{Phi-assoc}) takes the form
\be
    A \, T(B,C) - T(A B,C) + T(A,B C) - T(A,B) \, C = 0 \, ,  \label{T-cob}
\ee
where
\be
    T(A,B) := N( N(A) B + A N(B) - N(A B) ) - N(A) \, N(B)
\ee
is the Nijenhuis torsion of $N$. Thus $T$ has to be a coboundary. (\ref{T-cob})
in turn is solved if $T$ is a cocycle, i.e., if
there is a linear mapping $f : \A \rightarrow \A$ such that
\be
    T(A,B) = A f(B) - f(A B) + f(A) B  \; .
\ee
A simple solution is given by $f=0$, which implies the
\emph{associative Nijenhuis relation}
(see also \cite{Fuch97,CGM00,Ebra04}, for example)
\be
    N(A) \, N(B) = N( A N(B) + N(A) B - N(A B))  \; .  \label{Nijenhuis}
\ee
\vskip.2cm

\noindent
{\it Example 1.} Let us fix an element $P \in \A$. Then
\be
    N(A) = P \, A
\ee
solves (\ref{Nijenhuis}) and we have
\be
    \Phi(A,B) = A P B \; .
\ee
\hfill $\blacksquare$
\bigskip

\noindent
{\it Example 2.}
Let $A \mapsto \hat{A}$ be an involution of $\A$, so that
$\widehat{A B} = \hat{A} \, \hat{B}$. Setting
\be
    N(A) := {1 \over 2} A^{(-)} \qquad
    \mbox{where} \quad A^{(\pm)} := {1 \over 2} (A \pm \hat{A}) \, ,
\ee
$T$ is a cocycle with $f(A)= - (1/8) \, A^{(-)}$ and
\be
    \Phi(A,B) = A^{(-)} B^{(-)} \; .
\ee
Note that the involution leads to a direct sum decomposition 
$\A = \A^{(+)} \oplus \A^{(-)}$ where $A^{(+)}$ is a subalgebra 
and   $\A^{(+)} \A^{(-)} \subset \A^{(-)}$,  $\A^{(-)} \A^{(+)} \subset \A^{(-)}$,
and  $\A^{(-)} \A^{(-)} \subset \A^{(+)}$. 
In particular, every $\mathbb{Z}_2$-graded algebra has an obvious involution (see also 
\cite{Golu+Soko00} for the Lie algebra case).
This includes the important cases of Clifford and super-algebras.  
\hfill $\blacksquare$

\section*{Appendix B: `Undeforming' the product}
\addcontentsline{toc}{section}{\numberline{}Appendix B: 
`Undeforming' the product}
\setcounter{section}{2}
\setcounter{equation}{0}
Let us recall the definition of the product used in
section~\ref{sssection:CLL} (see also example~1 in appendix~A),
\be
    A \bullet B = A \, P \, B + \zeta \, A \, (I-P) \, B 
\ee
where $I$ is the unit matrix and $P^2 = P$. 
By a direct calculation one verifies that this product satisfies
\be
      Q (A \bullet B) Q = (Q A Q) (Q B Q)
\ee 
with
\be
    Q := P + z \, (I - P) \, , \qquad \zeta = z^2 \; .
\ee
For example, for $2 \times 2$ matrices 
\be
    A = \left(\begin{array}{cc} a & b \\ c & d \end{array}\right) \, ,
        \qquad
    P = \left(\begin{array}{cc} 1 & 0 \\ 0 & 0 \end{array}\right) \, ,
\ee
we have
\be
    Q = \left(\begin{array}{cc} 1 & 0 \\ 0 & z \end{array}\right) \, , \qquad
    Q A Q = \left(\begin{array}{cc} a & z \, b \\ z \, c & z^2 \, d
      \end{array}\right) \; .     \label{QAQ}
\ee
With $V$ given by (\ref{V-expansion}) we associate
\be
    \hat{V} := Q V Q  =  v_0 + v_1 \, z^{-2} + v_2 \, z^{-4} + \cdots
\ee
where the coefficients $v_n = Q u_n Q$ now exhibit an explicit $z$-dependence. 
The hierarchy equations 
\be
    V_{t_n} = (\zeta^n V)_{\geq 0} \bullet V - V \bullet (\zeta^n V)_{\geq 0}
\ee
can then be written in terms of $\hat{V}$ as 
\be
    \hat{V}_{t_n} = (z^{2n} \hat{V})_{\geq 0} \hat{V}
                    -\hat{V} (z^{2n} \hat{V})_{\geq 0} \, , 
\ee
if we agree upon the rule that the projection $(\, )_{\geq 0}$ does \emph{not} 
take the `inner' $z$-dependence of the $v_n$ into account (i.e. the $z$-dependence 
shown in the last matrix in (\ref{QAQ})). In this way contact is made with  
previous formulations of hierarchies of the type considered in this work 
(see, in particular, \cite{Veks98,Veks02}). By use of deformed products one achieves 
a much more elegant formulation, according to our opinion (and remains in the
class of `Gelfand-Dickey-type' hierarchies).

\end{document}